\documentclass{article}
\usepackage[utf8]{inputenc}
\usepackage{authblk}
\usepackage{color}
\usepackage{booktabs}
\usepackage{graphicx}
\usepackage{amsmath}
\usepackage{amssymb}
\usepackage{soul}

\newcommand{\blue}[1]{\textcolor{blue}{#1}}

\newcommand{\average}[1]{\langle #1 \rangle}
\newcommand{\numnodes}[0]{V}  
\newcommand{\popsize}[0]{N}   
\newcommand{\nimax}[0]{n^{\text{max}}}  
\newcommand{\ptravel}[0]{W}   
\newcommand{\neighbors}[0]{\mathcal{N}}
\newcommand{\ttransient}[0]{t_{\text{tr}}}
\newcommand{\ninfrvar}{M_{i}}
\newcommand{\ninfrvarj}{M_{j}}
\newcommand{\ninfval}{m_{i}}

\title{A Markov chain for metapopulations of small sizes with attraction landscape
}

\author[1]{Paulo C. Ventura}
\author[1]{Eric K. Tokuda}
\author[1]{Luciano da F. Costa}
\author[1]{Francisco A. Rodrigues}
\affil[1]{University of São Paulo, São Carlos, Brazil}

\date{}

\begin{document}

\maketitle

\begin{abstract}
Mathematical models represent one of the fundamental ways of studying nature. In special, epidemic models have shown to be particularly useful in the understanding of the course of  diseases and in the planning effective control policies. A particular type of epidemic model considers the individuals divided into populations. When studied in graphs, it is already known that the graph topology can play an important role in the evolution of the disease. At the same time, one may want to study the effect of the presence of an underlying \emph{attraction landscape} of the vertices, apart from the respectively underlying topology. In this work, we study metapopulations with small number of individuals in the presence of an attraction landscape. Individuals move across populations and get infected according to the SIS compartmental model. By using a Markov chain approach, we provide a numerical approximation to the prediction of the long-term prevalence of the disease. More specifically, an approach that combines two binomial distributions for mobility, with appropriate assumptions, is proposed to approximate the model. The problem setting is simulated through Monte-Carlo experiments and the obtained results are compared to the mathematic-analytical approach. Substantial agreement is observed between both approaches, which corroborates the effectiveness of the reported numerical approach. In addition, we also study the impact of different levels of attraction landscapes, as well as propagation on the local scale of the entire population. All in all, this study proposes a potentially effective approach to a mostly unexplored setting of disease transmission.
\end{abstract}

\section{Introduction} \label{sec:intro}

The history of humankind has been marked by several infectious disease outbreaks. Recently, the dynamical modelling of epidemics proved to be an important asset for the planning of intervention strategies~\cite{anderson1992infectious}. Their value lies within its ability to study potential future epidemic scenarios and analyze public health intervention possibilities~\cite{balcan2009seasonal,brown2004asian,aleta2020modelling}.


While epidemic models can be complex and incorporate arbitrarily high levels of realism, intricate systems often have features that are reasonably governed by simple rules. The class of compartmental models considers that, at a particular moment in time, each individual has a unique disease status~\cite{daley2001epidemic}. For instance, the classical SIR model considers three statuses: susceptible, infectious and recovered.  Therefore, a healthy individual, susceptible to the disease, could transition to the infectious state, and then to the recovered and immune states. There are variants of these states, such as the SIS, in which there are only two states and the individual never get completely immune to the disease~\cite{keeling2011modeling}.

A common assumption of compartmental and other epidemic models is the homogeneous mixing setting, which can be simply handled with a mean field theory~\cite{anderson1992infectious}. Alternatively, previous works considered heterogeneity in the contacts by considering metapopulations~\cite{hanski2000metapopulation,ball1991dynamic,cross2005duelling}. This formulation allows approaching several problems, such as the synchronization among populations~\cite{lloyd1996spatial} and the global epidemic thresholds~\cite{ball1997epidemics,cross2005duelling}.


A particular case of metapopulation consists in considering \emph{subpopulations} of small sizes, as observed in a network of households. The mathematical modelling of an epidemics contemplating a household structure was studied, for example, in~\cite{ball1999stochastic}. The population was partitioned into a large number of houses with a small number of individuals living in each house. The contagion occur in a two-level mixing, inside and outside the houses. They determine, under the described constraint, the invasion threshold considering a deterministic and a stochastic model.


In \cite{neal2006stochastic}, the authors studied the spread of SIS epidemics in a community of households using Markov chains and Markov birth-death processes. They derived a weak law or large numbers for the average number of infectives per household. For an exponential distribution of infectious periods, they also proved the existence of an endemic equilibrium. Furthermore, they established bounds for a stochastic Gaussian process.

Most previous works on metapopulation epidemic models (e.g.~\cite{colizza2007reaction,colizza2008epidemic,ajelli2010comparing}) considered a large number of individuals in each community. Works such as~\cite{ball1999stochastic,neal2006stochastic} assessed smaller-sized populations, but with simple mobility rules, with equal probability of reaching any of the neighbours. In both fields, most models were of the SIR class (this is, with a single outbreak), while a better understanding of the SIS class models for endemic diseases in metapopulations is recognized as a current challenge~\cite{ball2015seven}. 

No previous work seems to have considered an SIS model of a disease over a network of households with the migration of the individuals stochastically conditioned by a probability matrix. This incorporates the complementary effect of an arbitrary \emph{bias} in the mobility regardless of the network topology, i.e., nodes with attraction levels, independently of the topological features. 

We propose a Markovian approximation to the problem and compare it with a Monte-Carlo agent-based simulation. Two stages are taken into account, epidemic and mobility, in which agents can respectively change their epidemic state and their location state. We formulate both stages as Markovian dynamics. The epidemic stage is designed considering the combination of binomial distributions. The mobility stage in turn is modelled considering a combination of a binomial and a Poisson distribution which, for the implementation, is approximated to two binomials. For the parameters analyzed, we observed considerable agreement of the proposed model with the asymptotic behavior of the simulations. In particular, the behavior of the stationary disease prevalence with the transmission probability $\beta$ and the steepness of the attraction was analysed. We also checked the agreement between Markovian and Monte-Carlo formulations in a node level. 

In the next section, the model studied in this work is described. In Section~\ref{sec:markov}, we devise a numerical description of the 
dynamics of the considered model. In Section~\ref{sec:experiments} we describe the experimental setup and discuss the main results. And finally, in Section~\ref{sec:conclusion}, we present our conclusions as well as future works.

\section{Model} \label{sec:model}

\begin{figure}[ht]
  \centering
 \includegraphics[width=0.8\textwidth]{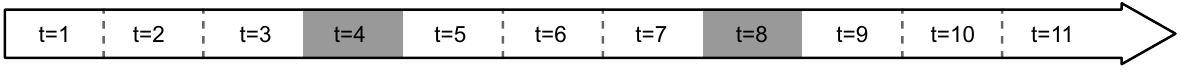}  
   \caption{Epidemic and mobility steps in the simulation timeline. In this example of $r=4$, each mobility step (shaded boxes) happens after three epidemic steps.}
  \label{fig:diagsteps}
\end{figure}

In this section we describe how we model the epidemic process, the transmission of disease and how the individuals move. We approach this problem through a discrete-event system. A fundamental assumption of this model is that \emph{events} are responsible for changing the system state. The events occur at discrete time steps and the system state is not allowed to change between subsequent time steps. Monte Carlo simulations additionally require the existence of a (pseudo-) random number generator. In this work, each time step contains at least one of the two stages: epidemic and mobility. In the standard scenario, every time step is composed of both steps, in a sequential way. Alternatively, by defining an epidemic-mobility ratio $r$, one of the phases may be suppressed in the time step. For instance, for $r=2$, the mobility phases every second step, i.e., interleaved with the epidemic steps. Analogously, for $r=4$, the agents move after three epidemics steps (see Fig.~\ref{fig:diagsteps}).

The epidemic model considered is the SIS compartmental model \cite{keeling2011modeling}. The main reason for this choice is the intrinsic simplicity of this model, due to the presence of just two possible epidemic states for each individual. Every  healthy (susceptible) individual in contact with an infectious individual may get infected with probability $\beta$. The infectious individual, in turn, may heal itself with probability $\gamma$.

\begin{table}[ht]
\centering
\caption{Notation.}\vspace{.5em}
\label{tab:notation}
\begin{tabular}{ll}
\toprule
Notation & Description \\
\midrule
$\numnodes$	& Number of nodes \\
$\popsize$	& Total number of individuals  \\
$N_i$      	& Random variable of the number of individuals in node i  \\
$\ninfrvar$       & Random variable of the number of infectious individuals in node i  \\
$M$         & Random variable of the number of infectious individuals  \\
$\rho_I$    & Total prevalence (fraction of infectious individuals) \\
$\beta$		& Individual contact transmission probability \\
$\gamma$	& Individual healing probability \\
$t$       	& Step in the discrete simulation    \\
$r$         	& Epidemic-mobility ratio\\
$a_i$         	& Attraction level of node $i$\\
$\xi$		& Exponent of the attraction levels\\
$W$		& Mobility stochastic matrix\\
\bottomrule
\end{tabular}
\end{table}

The epidemic dynamics takes place on a graph, in a metapopulation setting~\cite{masuda2010effects}. The epidemic propagation only occurs between agents located at a same node, inside which homogeneous mixing is assumed under the pseudo mass-action law~\cite{keeling2011modeling} (this is, all agents interact with each other during each time step, and the interaction strength does not decrease with the population size as in the mass-action law). Therefore, at each time step, the individuals at node $i$, initially with $m_i$ infectious individuals, have a chance $1 - (1 - \beta)^{M_i}$ of catching the disease and becoming infectious. At a given instant of time, every individual occupies a node and can go to another neighbouring node. Differently from most of metapopulation models, we focus on populations of small sizes. As mentioned in the introduction, most works that consider this feature refer to networks of households\cite{ball1997epidemics,ghoshal2004sis,house2008deterministic}, in which individual mobility is not considered in an agent-based scheme.

\begin{figure}[ht]
  \centering
 \includegraphics[width=0.5\textwidth]{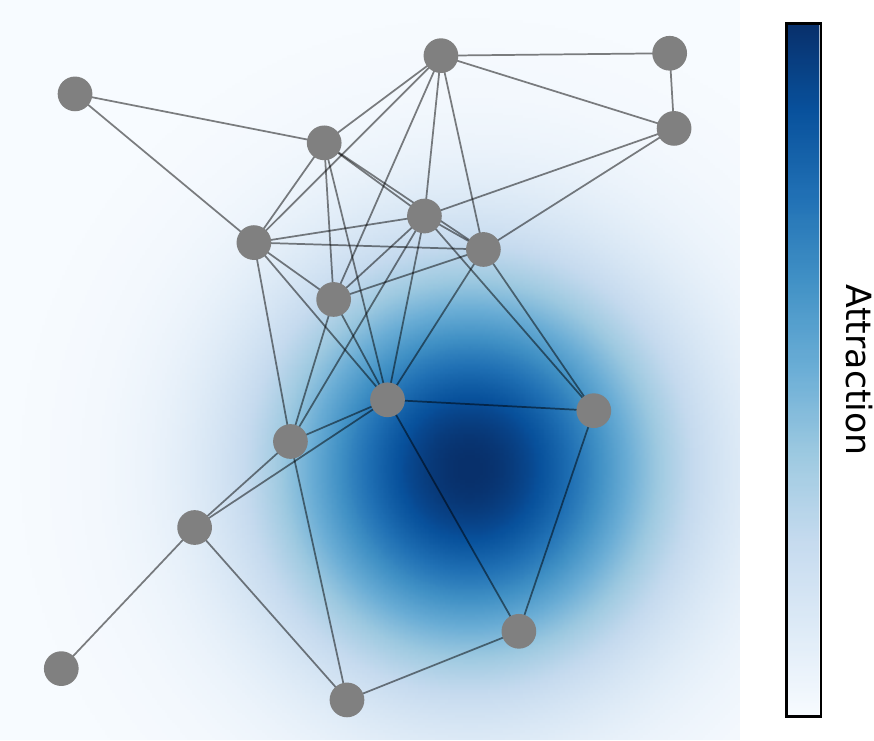}  
   \caption{A graph overlaid on top of an attraction landscape. The landscape associates an attraction value $a_i$ for each node $i$, given by the position of $i$. The mobility of individuals is stochastically guided by these values.}
  \label{fig:gradient}
\end{figure}

The motion of the individuals occurs stochastically as a random walk process. Each move takes into account the attraction landscape, embedded in space. Every node $i$ with a given set of neighbouring nodes $\neighbors_i$ (which, by construction, includes $i$ as a self-loop) associates to a value $a_i$ in the map. Individuals in $i$ move to neighbours with probability proportional to their attraction values. More precisely, the probability of an individual in $i$ to move to the neighbouring node $j \in J$, $p(i\rightarrow j)$, is given by: 

\begin{equation}
    \label{eq:ptravel_def}
    p(i\rightarrow j)= \ptravel_{ij} = a_j / \sum_{k \in \neighbors_i} (a_k). 
\end{equation}

Such probability values do not change over time and the matrix organization corresponds to the stochastic matrix of this mobility dynamics. Ultimately, the movement of the individuals in the graph is a simple random walk with the effect of the attraction map fully encoded in the obtained stochastic matrix.

\section{Numerical approximation} \label{sec:markov}

Epidemic models in populations with household structure have been extensively studied with mathematical formulations in previous works~\cite{ball1997epidemics,ball1999stochastic,ghoshal2004sis,fraser2007estimating,house2008deterministic}. As mentioned, these works simplify the host mobility by assuming that individuals from different households interact weakly, but in general homogeneously. On the other hand, more works with epidemics in metapopulations focus on the case of many individuals per node, which is suitable to the study of cities. In this case, the epidemic dynamics can be well approximated by the temporal evolution of the average numbers of individuals in each state at each site, disregarding the unfeasible space of each probability. In our work, we develop a numerical approximation for the case of small populations with individual-level mobility.

The approximation that we propose is based on a discrete-time Markov chain. While the dynamic of the epidemic step inside a node is simple and can be modeled exactly for a finite population, the exact Markov chain for the mobility step is particularly complex. We explain how we address this problem by using an appropriate distribution to approximate the outcomes of the mobility step.

\subsection{Epidemic step}
\label{sec:markov_epidemic_step}

A sequence of consecutive epidemic steps inside a given node $i$ can be described by the joint probabilities in which, at step $t$, the node is populated by $n_i$ individuals, out of which $\ninfval$ are infectious, given by $p\big(\popsize_i(t)=n_i, \, \ninfrvar(t) = \ninfval \big)$. During this process, however, the total number of individuals $N_I$ does not change, meaning that the transition matrix of the whole Markov process is block-diagonal with respect to $n_i$. For a given occupation $\popsize_{i}(t) = n_i$ and infectious count $\ninfrvar(t) = \ninfval$, each susceptible agent can become infectious with probability $1 - (1 - \beta)^{m_i}$, while each infectious agent can remain infectious with probability $1 - \gamma$. The number of infectious agents in the next step $\ninfrvar(t)$ is thus a sum of two binomial random numbers: the infection of $n_i - \ninfval$ susceptible agents and the ``non-healing'' of $\ninfval$ infectious ones. The transition matrix element for this Markov process can thus be written as:

\begin{align}
    p\big(\popsize_{i}(t+1) = n_{i}', \, \ninfrvar(t+1) = \ninfval' \; \big| \; \popsize_{i}(t) = n_{i}, \, \ninfrvar(t) = \ninfval\big) \; =  \nonumber \\
    = \delta_{n_i, n_i'} \sum_{m = 0}^{n - n_{i, I}} \binom{n_i - \ninfval}{m} [1 - (1 - \beta)^{\ninfval}]^m \; (1 - \beta) ^{(n_i - \ninfval) - m} \;\cdot \nonumber \\ 
    \cdot \binom{\ninfval}{\ninfval' - m}  (1 - \mu)^{\ninfval' - m} \cdot \mu^{\ninfval - (\ninfval' - m)}
    \label{eq:epid_transmat}
\end{align}

\begin{figure}[ht]
	\centering
	\includegraphics[width=0.75\textwidth]{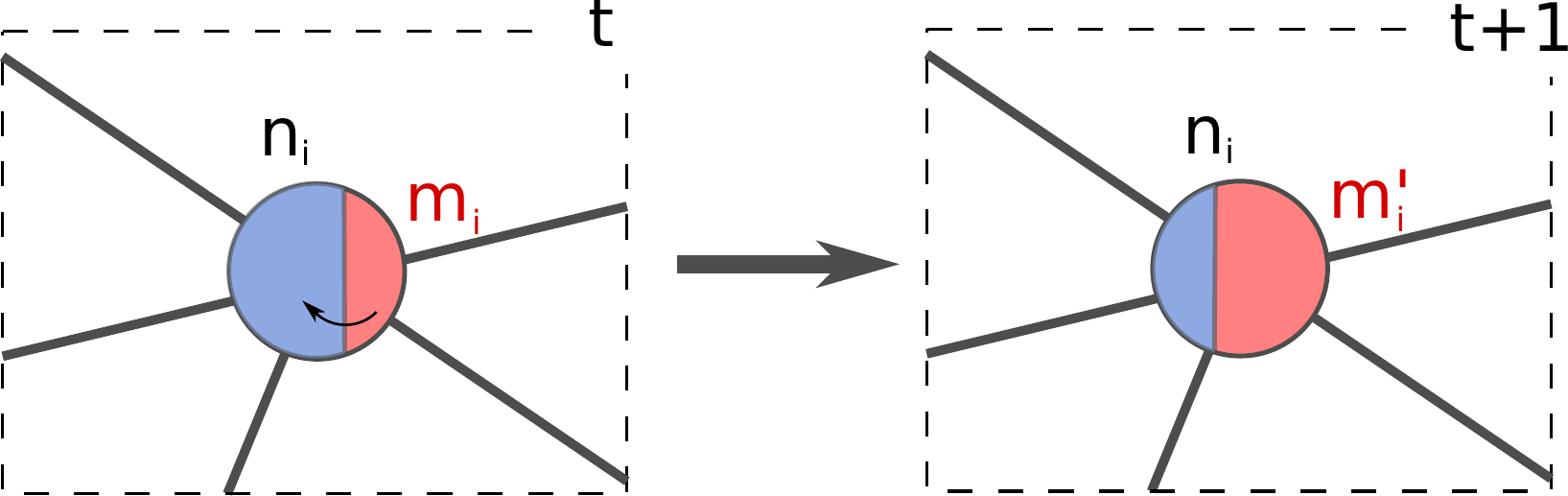}
	\caption{Epidemic step on node $i$. In the homogeneous mixing setting, individuals in the same node interact with each other and the number of infectious $m_i$ may change ($m_i'$). Individuals do not move between nodes in this step and, as a consequence, the number of individuals per node $n_i$ remains unaltered.}
	\label{fig:stepepid}
\end{figure}

where the mute index $m$ represents the number of infections during the current step. If we assume a maximum number of individuals $\nimax_i$ in node $i$, for which $p(\popsize_{i} > \nimax_i) = 0$ at any time step, the epidemic Markov chain can be numerically propagated with complexity $\mathcal{O}((\nimax_i)^2)$.
For multiple consecutive epidemic steps, it is enough to calculate the powers of the transition matrix just once and reuse them during the calculations, therefore without increase of the computational complexity.

\subsection{Mobility step}
For the mobility stage, however, the exact problem has a substantially greater complexity even for a single step. The system must be, in principle, represented by joint probabilities of the whole system, considering all possible values of $\popsize_{i}$ and $\ninfrvar$ for each node $i$. One first and usual approximation~\cite{van2008virus,gomez2010discrete,pastor2015epidemic} is to neglect the dynamical correlations between the variables of different nodes, which allows the factorization of the whole system into the node-wise dynamics of $p\big(\popsize_{i}(t) = n_i, \ninfrvar(t) = \ninfval\big)$. One consequence is that the total number of agents $\popsize$ is a random variable and could, in principle, fluctuate. We need to ensure that at least $\average{\popsize}$ is constant in time for physical consistency. 

\begin{figure}[ht]
	\centering
	\includegraphics[width=0.75\textwidth]{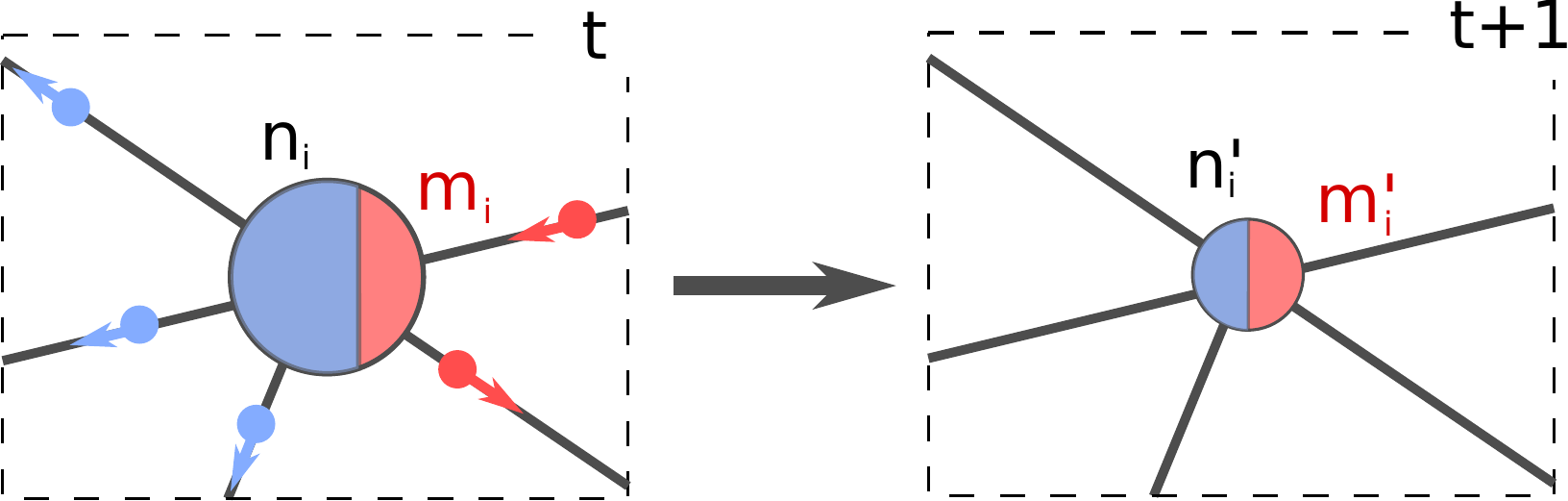}
	\caption{Mobility step on node $i$. In this step, while each individual maintains its infectious status, they are allowed to move to neighbouring nodes. As a result, both the number of individuals $n_i$ and the number of infectious $m_i$ in the node may change. }
	\label{fig:stepmob}
\end{figure}

In our setting, as we are considering the microscopic occupation numbers of the nodes, the problem becomes prohibitively complex. For example, the total number of agents in node $i$ after a random walk step depends on each possible number of agents in its immediate neighbors, including itself, combined with each possible number of travelers coming from each neighbor. The number of combinations is further increased when we distinguish susceptible from infectious agents.

To address this issue, we assume $p(\popsize_{i} = n_i, \ninfrvar = \ninfval)$ after a mobility step based on less information than the full probability distributions. Consider the expected occupations after the mobility step:



\begin{align}
    a_i := \average{\popsize_i(t+1)} = \sum_{j \in \neighbors_i} \average{\popsize_{j}(t)} \cdot \ptravel_{ji} \label{eq:avg_next_n_mobility}\\
    b_i := \average{\ninfrvar(t+1)} = \sum_{j \in \neighbors_i} \average{\ninfrvarj(t)} \cdot \ptravel_{ji} . \label{eq:avg_next_ni_mobility}
\end{align}

Eqs.~\ref{eq:avg_next_n_mobility} and \ref{eq:avg_next_ni_mobility} consider the contribution of each neighbor of $i$ to its own average agent occupation after the mobility step and, in the context of neglecting dynamical neighboring correlations, is rather exact. The approximation we propose is to use a reasonable distribution for $p\big(\popsize_{i}(t+1) = n_i, \ninfrvar(t+1) = \ninfval\big)$ based on $a_i$ and $b_i$. For example, as we stated, there is typically a very high number of ways in which the number of agents in $i$ after a mobility step reach $n_i$, each of these ways having a very small probability of occurring. If all these probabilities were the same, then $n_i$ would be a binomially distributed variable, with good approximation to a Poisson distribution, as the number of events is typically very high. We can consider this as a reasonable guess for the distribution of $n_i$. 
A similar hypothesis can be made for the number of infectious agents $\ninfval$, though in this case it is upper bounded by $n_i$, which is typically not too large and therefore is better modeled with a binomial distribution. We can state that each agent that arrives at node $i$ has a chance $b_i / a_i$ of being infectious.

Under this hypothesis, the distribution of $n_i$ and $\ninfval$ just after a mobility update is approximated as an independent combination Poisson and a Binomial distributions:


\begin{align}
    p\big(\popsize_{i}(t+1) = n_i\big) \;&= e^{-a_i} \frac{a_i^{n_i}}{n_i!} 
        \label{eq:binomial-poisson_pn}\\ 
    p\big(\ninfrvar(t+1) = \ninfval\, | \, \popsize_{i}(t+1) = n_i \big ) \;&=  \binom{n_i}{\ninfval} \left( \frac{b_i}{a_i} \right)^{\ninfval} \left( 1 - \frac{b_i}{a_i} \right)^{n_i - \ninfval} .
    \label{eq:binomial-poisson_pm}
\end{align}

The final joint probability $p\big(\ninfrvar(t+1) = \ninfval , \, \popsize_{i}(t+1) = n_i \big )$ is then the product between the right-hand sides of Eqs. \ref{eq:binomial-poisson_pn} and \ref{eq:binomial-poisson_pm}. Along with the epidemic transition matrix given by Eq. \ref{eq:epid_transmat}, Eqs. \ref{eq:binomial-poisson_pn} and \ref{eq:binomial-poisson_pm} completely define a dynamical system. It can be analytically shown that the proposed mobility step, given by Eqs.~\ref{eq:avg_next_n_mobility} to \ref{eq:binomial-poisson_pm}, preserves the average numbers of agents $\average{N}$ and infectious agents $\average{M}$ for the entire metapopulation, which is physically consistent. 

However, it is not feasible to consider the entire Poisson distribution for $n_i$, which ranges from $0$ to $+\infty$ in the numerical evaluation. When the number of agents is smaller than or comparable to the number of subpopulations, we can truncate the distribution up to a reasonably chosen number $\nimax_i$ for each node $i$, so that $n \leq \nimax_i $ comprises most of the cumulative occupation probability. However, simply truncating the Poisson distribution breaks the normalization of the distribution, i.e., the condition that $\sum_{n_i = 0}^{\nimax_i} p(\popsize_i = n_i) = 1$. Even if a forced normalization is applied, dividing the truncated probabilities by their sum, the average population after the mobility step is not preserved (this is, $\average{N(t+1)} \neq \average{N(t)}$).

As an alternative, we instead replace the Poisson distribution by a Binomial one with $\nimax_i$ attempts and equal average and, given by:

\begin{equation}
    p\big(\popsize_{i}(t+1) = n_{i}\big) \;= \binom{\nimax_i}{n_i} \left( \frac{a_i}{\nimax_i} \right)^{n_i} \left( 1 - \frac{a_i}{\nimax_i} \right)^{\nimax_i - n_i}
    \label{eq:binom_approx}
\end{equation}

with an average $a_i$, as required. This approximation relies on the similarity between the Poisson an Binomial distribution with equal averages and sufficiently high $\nimax_i$. The proposed Binomial distribution is normalized and can also be shown to maintain the overall average population $\average{N}$, which justifies the choice.


\section{Experiments and discussion}
\label{sec:experiments}

In this section we present the simulation setup and discuss the experimental results.

Previous works on usual metapopulation settings often evaluate the influence of different network topologies on the epidemic course. Instead, in this work we focus on a single topology, the random geometric graph (RGG), so we can isolate the effect of the complexity layer added by the particularities of the problem proposed, namely the small sizes of populations and the presence of an attraction landscape.

In the RGG model, nodes are randomly placed in the space following a uniform probability density distribution. A central parameter of this model is $d$: nodes closer than $d$ are connected. This original formulation generates an undirected network.

\subsection{Simulation}

The Monte-Carlo simulation was implemented using an Agent-based model (ABM). ABM is a stochastic simulation paradigm focused on basic building blocks, the \emph{agents}~\cite{macal2016everything}. An agent is governed by simple rules and one of the central interest in using ABMs is related to the exploration and understanding of the collective behavior of them. 
Here, the agents correspond to the individuals. Each agent \emph{state} is described by its epidemic state (susceptible or infectious) and its location (node). The system state is described by the state of all agents. Every simulation step may comprise two phases: an epidemic step and a mobility step. 

One may consider the evaluation of consecutive epidemic (or mobility) phases by changing the simulation parameter epidemic-mobility ratio ($r$).
The epidemic and mobility phases follow the model described in Section~\ref{sec:model}. In particular, they heavily depend on the simulation parameters $\beta$ and $\gamma$ and the stochastic matrix. Multiple agents can occupy the same vertex. The simulation starts with a sample of the agents already infectious and the dynamics is analyzed for a fixed time span. The (pseudo-) random generation of numbers is a fundamental piece and, here, we store the seed for reproducibility.

All experiments, except when mentioned, considered the parameter values shown in Table~\ref{tab:simuparams}. 1000 executions were performed for each point, each with a different network sample. During each execution, which is performed for 1000 steps (which are either mobility or epidemic steps), the first 800 steps are discarded, and the remaining 200 are used to calculate the steady-state value of the prevalence. In particular, the number of agents is equal to the number of vertices, which is a fundamental characteristic of this metapopulation setting: the low number of individuals per site. The RGG topology, as explained, was chosen because it provides a natural embedding in space, where the attraction landscape is also defined.

\begin{table}[ht]
\centering
	\caption{Parameters of the simulation}\vspace{.5em}
\label{tab:simuparams}
\begin{tabular}{ll}
\toprule
Graph topology		& RGG   \\
Number of vertices	& 1024  \\
Average degree		& 5  \\
Number of agents	& 1024 \\
Nr. of agents per vertex ($\eta$) & 1 \\
Simulation period	& 1000 \\
Initial number of infectious & 102 \\
Individual probability of healing ($\gamma$)		& 0.1 \\
Number of realizations  & 1000\\
\bottomrule
\end{tabular}
\end{table}

The simulation parameters $\beta$, $r$ and $\xi$ play a pivotal role in this dynamics and their impact on the epidemic model is evaluated separately.

\subsection{Validation of the numerical analysis}

We start by checking the capability of the Markov chain approximation to predict the stationary prevalence of the experiments. For the Monte Carlo experiments, 1000 executions are performed for each point, each with a different network sample. For the Markov chain numerical approach, we promote the epidemic and mobility iterations until the values of $\average{n_i}$ and $\average{m_i}$ do not change more than $10^{-9}$ in absolute value, for all nodes $i = 1, ..., V$. 

The average occupations $\average{n_i}$ of each node $i$ can be calculated in advance as $N \cdot x_i$, where $x_i$ is the leading eigenvector of the stochastic random walk matrix $W$ (defined in Eq.~\ref{eq:ptravel_def}), allowing us to initialize the process with these values. The calculated $\average{n_i}$ values are also used to reasonably define the truncation point of the node occupations as $\nimax_i = \max (5, \, 3\average{n_i})$. The minimum of 5 ensures that all nodes consider occupations from $n_i = 0$ to $5$, and nodes that are more visited are truncated at 3 times the average occupation, which comprises most statistically relevant states. We validated this choice by checking that higher values of $\nimax$ do not significantly change the results. 

\begin{figure}[ht]
    \centering
    \includegraphics[width=0.45\textwidth]{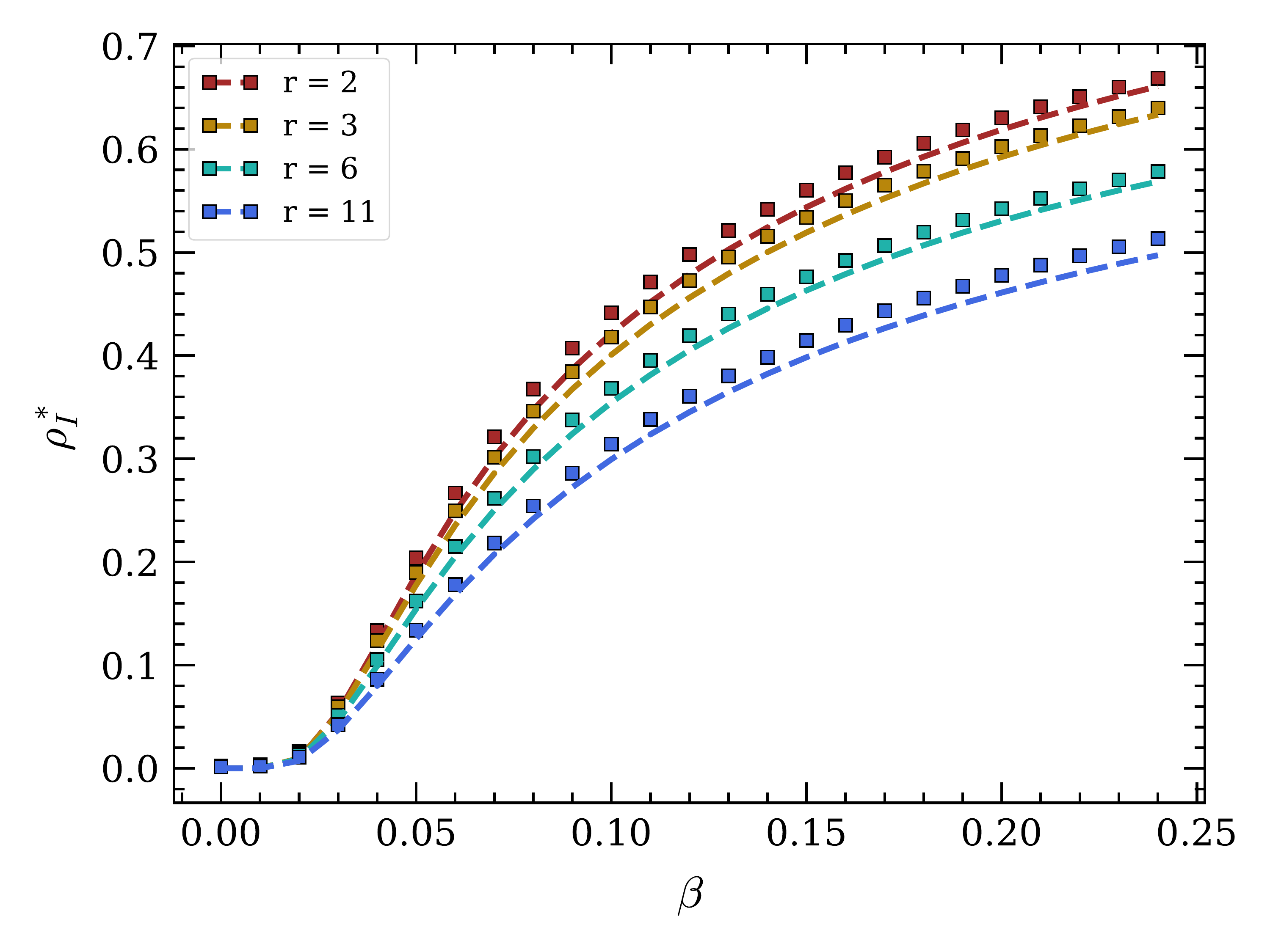}
    \includegraphics[width=0.45\textwidth]{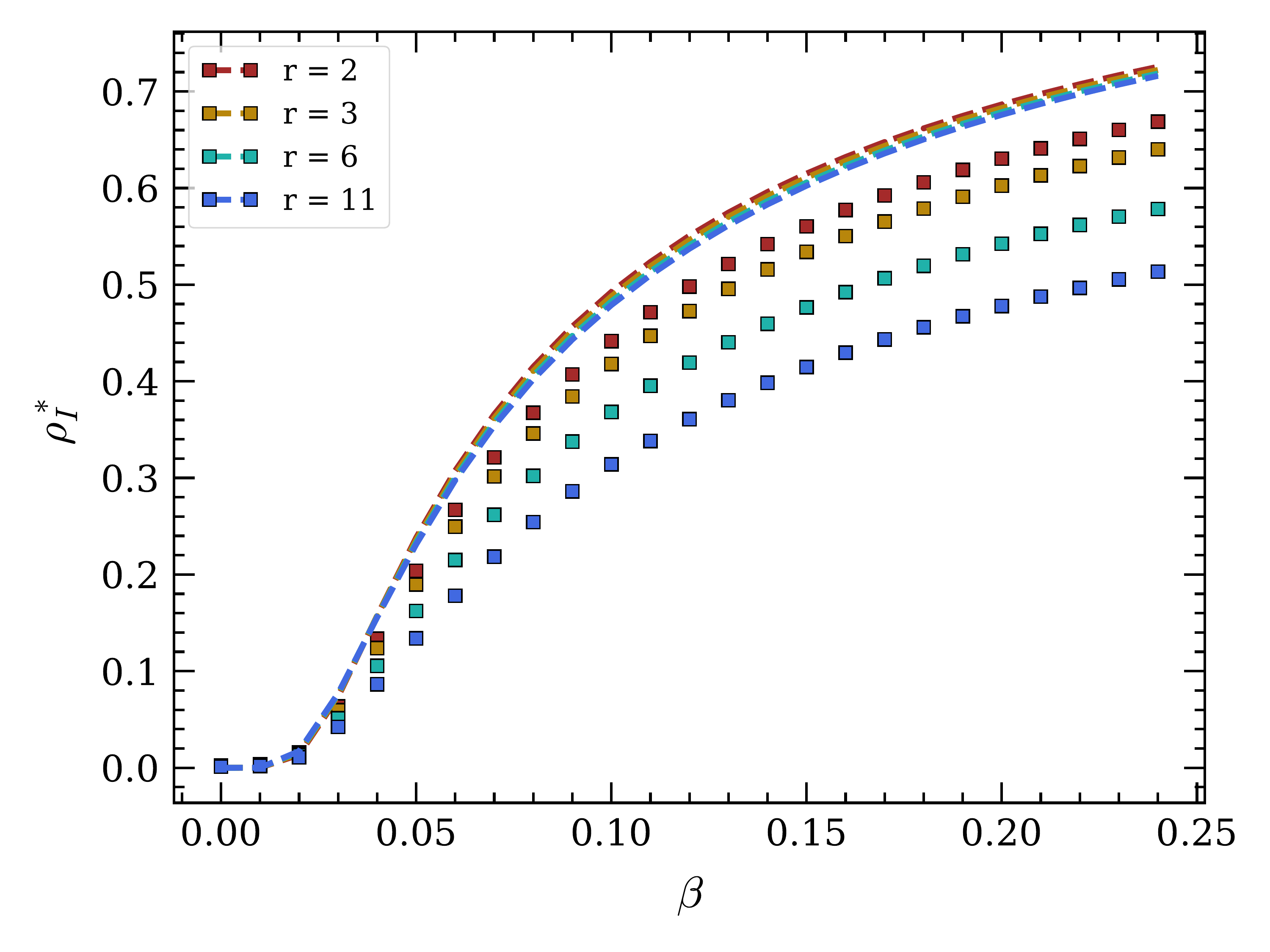}
    \caption{Stationary prevalence as a function of the individual transmission probability $\beta$, for the Monte Carlo simulations (square markers) and the numerical Markov chain  approach (dashed lines) using our proposed method (left) and the continuous population hypothesis (right).  Other parameters are set to: $\gamma = 0.10$, 
    $\eta = 1$, $\ttransient = 800$, $t_{\text{max}} = 1000$. Each point is an average over 1000 runs.
    }
    \label{fig:epidemic_curves}
\end{figure}

Fig.~\ref{fig:epidemic_curves} (left) shows the comparison between Monte Carlo experiments and numerical approximation, displaying the stationary prevalence $\rho_I^*$ as a function of the transmission probability $\beta$ for different values of the epidemic-mobility ratio $r$. The curves behave as expected, with higher prevalence observed for higher transmission. Notice that the transition between the healthy and endemic phases seems to occur smoothly. This is an intentional artifact, as each execution of the simulation generates a new network sample, each one presenting its own critical $\beta$ value for the transition. 

Another observed behavior is that greater epidemic-mobility ratios $r$ cause smaller stationary prevalences in the endemic phase. Higher $r$ means that more epidemic steps are performed before each mobility step, implying that the individuals are moving less, therefore justifying the smaller prevalence. It is commonsense, and shown for other metapopulation models~\cite{cross2005duelling}, that less mobility implies weaker disease spreading.

The agreement between the Monte Carlo simulations (square markers) and Markov chain numerical predictions (dashed lines) is remarkable. This shows that, even though the proposed Markov process requires deep approximation hypotheses, it is still able to capture the stationary prevalence of the simulations precisely. This happens specially because, as explained, the proposed Markov chain process accounts for the small number of individuals in each node $i$ by considering the probabilities of each occupation state $p\big(\popsize_i=n_i, \, \ninfrvar = \ninfval \big)$ for $n_i$ up to $\nimax_i$, instead of just the average values $\average{N_i}$ and $\average{\ninfrvar}$.

To verify this point, we test the results of a similar Markov chain formulation that disregards the probabilities and considers only the averages. This is done by changing the epidemic step, described in Section~\ref{sec:markov_epidemic_step}, into an iteration process over the quantities $\average{\ninfrvar(t)}$ according to the equation:

\begin{equation}
    \label{eq:contpop_epid_step}
    \average{\ninfrvar(t + 1)} = \average{\ninfrvar(t)} \cdot (1 - \mu) \,+\, (\average{N_i} - \average{\ninfrvar}) \cdot (1 - (1 - \beta)^{\average{\ninfrvar}}).
\end{equation}

Meanwhile, the mobility step is still simply performed by iterating $\average{\numnodes_i}$ and $\average{\ninfrvar}$ with Eqs. \ref{eq:avg_next_n_mobility} and \ref{eq:avg_next_ni_mobility}. Notice that there is no need to expand the averages $\average{\numnodes_i}$ and $\average{\ninfrvar}$ into a phenomenological probability distribution $p\big(\popsize_i=n_i, \, \ninfrvar = \ninfval \big)$ in this case; the process directly updates the averages. This method should approximate well the simulations if the number of individuals in each node were much higher than what we consider here, as already done in metapopulations with recurrent mobility \cite{gomez2018critical,shao2022epidemic} and random walks \cite{ball1991dynamic,hernandez2015discrete}.

The stationary prevalences as calculated by this alternative approach are shown as dashed lines in Fig.~\ref{fig:epidemic_curves} (right). Notice that this method notably overestimates the stationary prevalence, and almost does not capture the dependence with the epidemic-mobility ratio $r$. One reason for the overestimation is that this approach neglects that, when the number of individuals is low, the epidemic process has an appreciable probability of \emph{absorption}, this is, falling into the absorbing state $\average{\ninfrvar} = 0$. On the other hand, our Markov chain approach described in Section~\ref{sec:markov} provides numerically good estimates, even though it relies on a reductionist guess for the probability distributions.

\subsection{Intensity of the attraction} 

We now proceed to study the effect of the attraction landscape that is applied to the nodes to guide the agents' random walk. We can control the sharpness of the spatial attraction peak by raising the attraction intensity value of each node $i$  by a given exponent $\xi$ (this is, $a_i \rightarrow (a_i)^{\xi}$ for each $i = 1, ..., \numnodes$), before the directed edge weights $W_{ij}$ are calculated according to Eq.~\ref{eq:ptravel_def}. Provided that the original values of the attraction are no greater than $1$, this causes the differences between high-score and low-score nodes to be enhanced when $\xi > 1$, and diminished when $\xi < 1$. In other words, $\xi$ controls how heterogeneous is the attraction landscape.

\begin{figure}[ht]
    \centering
    \includegraphics[width=0.65\textwidth]{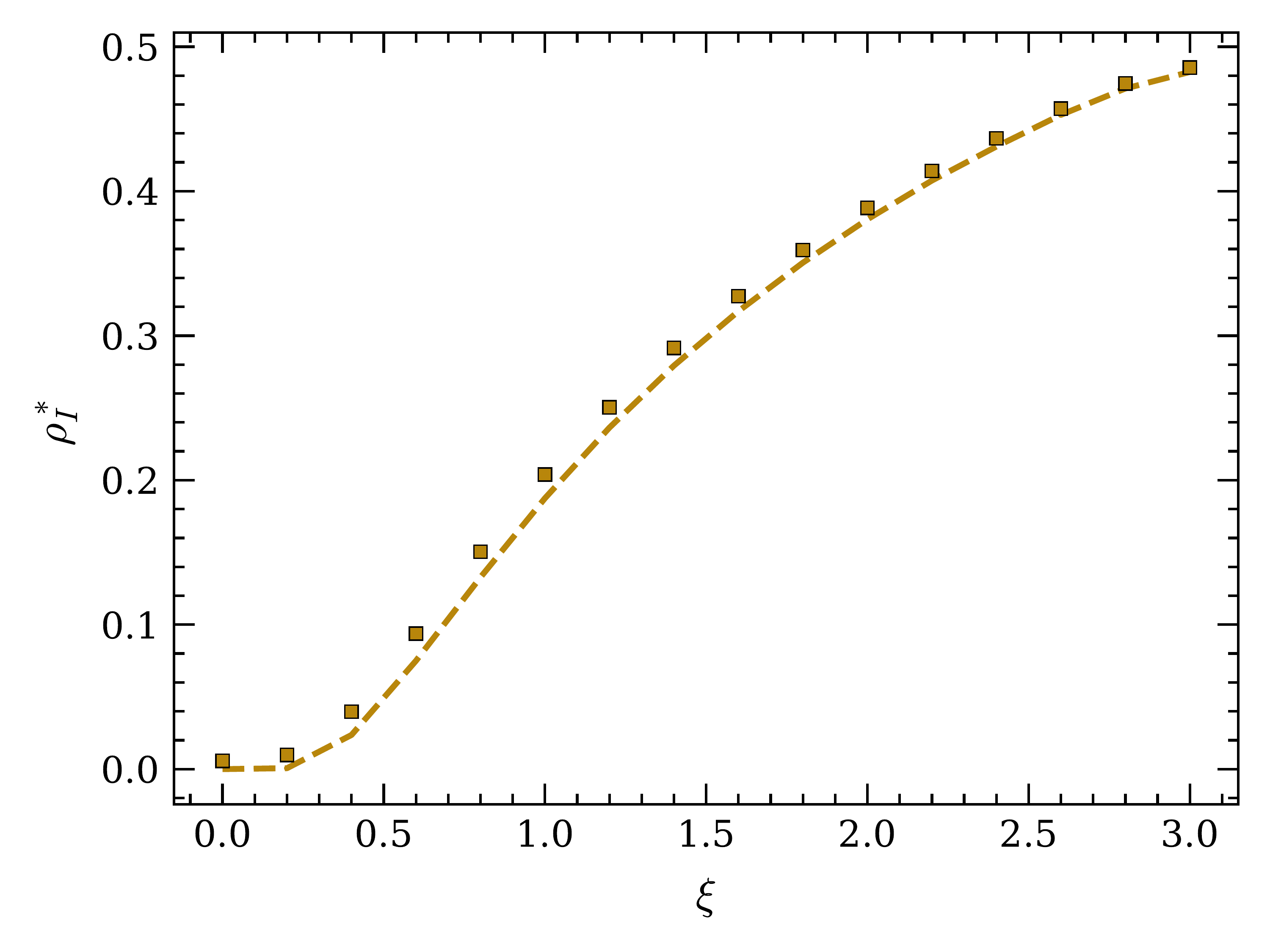}
    \caption{Stationary prevalence as a function of the exponent that is applied to the attraction values. 
    Parameters are set to $\beta = 0.05$, $r = 2$, $\gamma = 0.10$, $\eta = 1$, $\ttransient = 800$, $t_{\text{max}} = 1000$. Each point is an average over $1000$ runs.
    }
    \label{fig:gradexponent}
\end{figure}

In Fig.~\ref{fig:gradexponent}, we plot the stationary prevalence of the model for different values of the exponent $\xi$ on the same RGG topology ensemble for a fixed transmission probability $\beta = 0.25$. The sharpening of the attraction peak, achieved by increasing the exponent, $\xi$ causes an increase on the prevalence. This agrees with the fact that, in general, more heterogeneous topologies are better at propagating a disease, either in networks of individuals~\cite{pastor2001epidemic,barthelemy2005dynamical} or in networks of populations (metapopulations)~\cite{colizza2007reaction}. In our case, as we employ the pseudo mass-action law, this is even more accentuated: the landscape heterogeneity causes agents to gather around the highest score nodes, and the transmission events are more likely at nodes with more individuals.

Again, the Markov chain approach displayed marked agreement with the Monte Carlo simulations, capturing the correct behavior and offering numerical precision.

The increase of disease propagation with the heterogeneity of the metapopulation can be simply explained by the formation of agent gathering on highly visited nodes, which sustain the propagation. In the next section, we show this point by looking at the statistics of each node of the network.

\subsection{Local propagation metrics}

In order to study the disease propagation at a local scale, we perform Monte Carlo simulations with a single RGG network sample instead of a new one at each repetition, as in the previous section. Thus, we can determine the participation of each node into the epidemic and mobility dynamics. 

\begin{figure}
    \centering
    \includegraphics[width=0.48\textwidth]{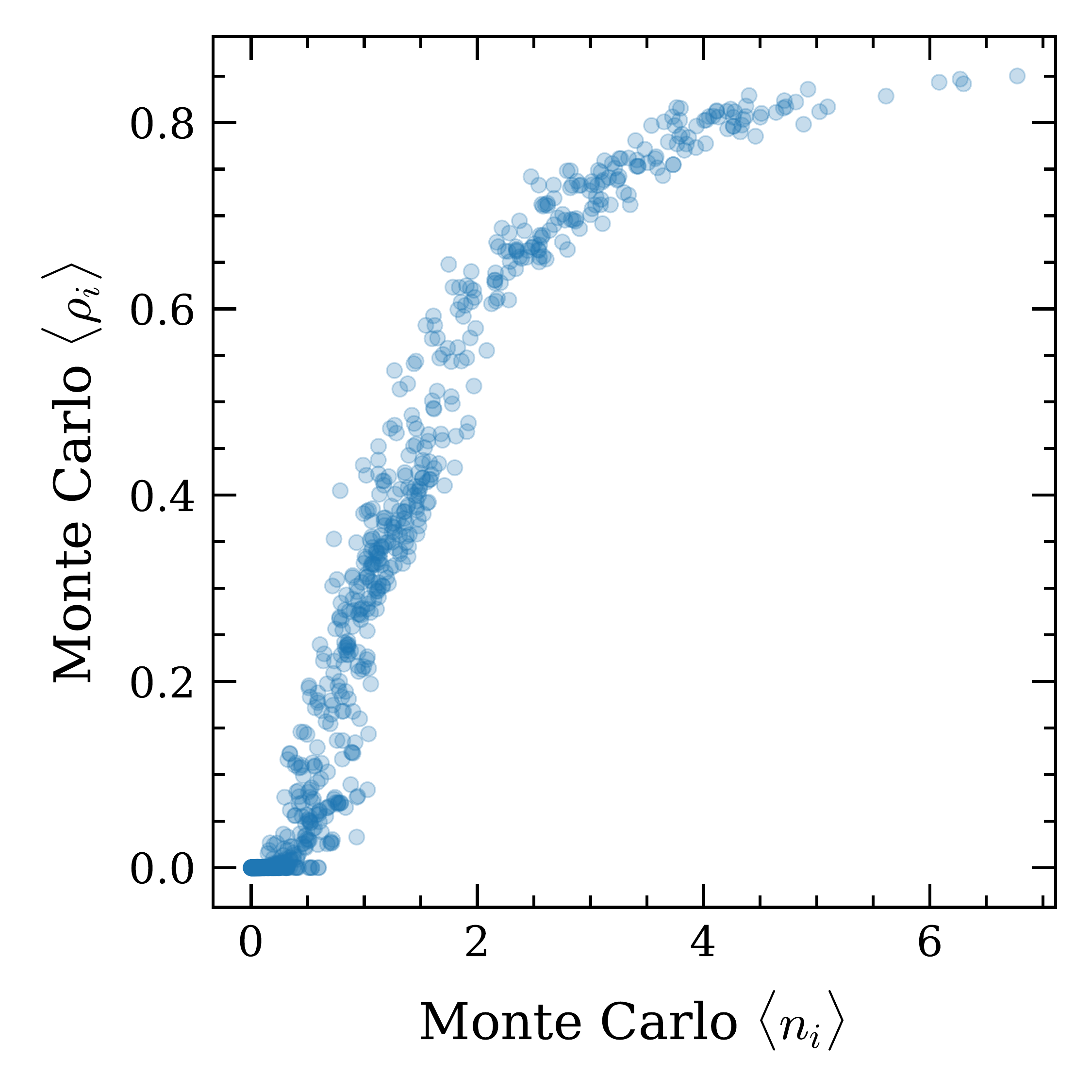}
    \includegraphics[width=0.48\textwidth]{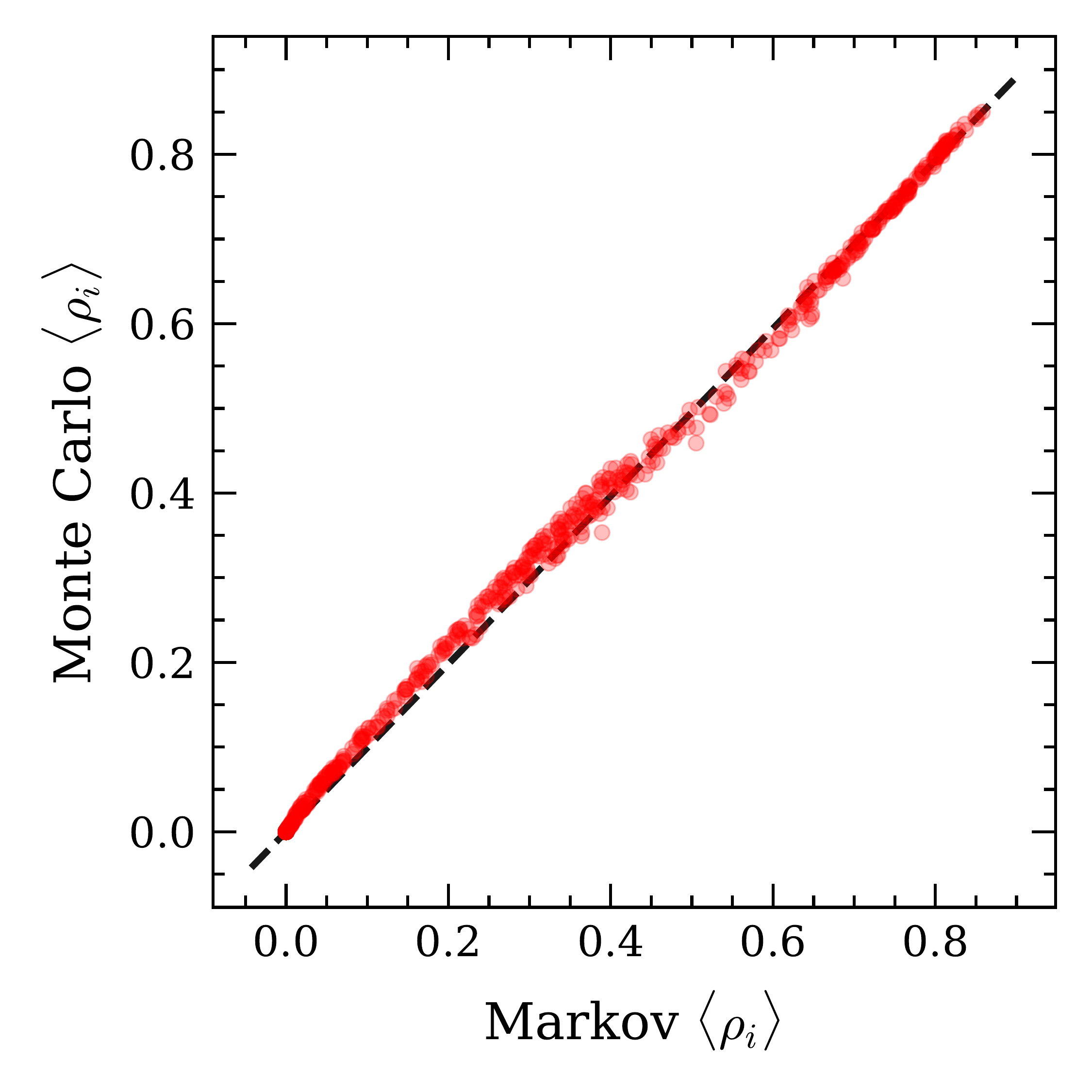}
    \caption{(Left) Average fraction of infectious individuals $\average{\rho_i}$ (disease prevalence) and the 
    average number of individuals $\average{n_i}$
    in each node $i$, represented as a point, as calculated from Monte Carlo simulations.
    (Right) Average fraction of infectious individuals as a function of the RW eigenvector entry for each node. Parameters are set to: $\beta = 0.2$, $r = 2$, $\gamma = 0.1$, $\eta = 1$, $\ttransient = 800$ and $t_{\text{max}} = 1000$. Each point is an average over 1000 independent runs.
    }
    \label{fig:scatters}
\end{figure}

In Fig.~\ref{fig:scatters}.a), we represent the average fraction of infectious agents $\average{\rho_i}$ (disease prevalence) and the average number of individuals $\average{n_i}$ at each node (represented as a point), averaged over time and the 1000 executions of the Monte Carlo simulation. At each time step, the prevalence $\rho_i$ at node $i$ is calculated as ${\ninfval} / n_i$, and defined as $0$ if $n_i = 0 $. The transmission probability $\beta = 0.2$ was chosen such that the disease is spread to a considerable fraction of the population, thus far above the epidemic threshold. 

The nodes that are more visited (represented by a higher value of $\average{n_i}$) sustain not only a higher \emph{number} of infectious agents, but also a higher \emph{fraction} of these, as evidenced in Fig.~\ref{fig:scatters}.a). This means that highly visited nodes are also hubs for the epidemic transmission. Because we use the pseudo-mass action law in each node, a higher number of individuals sharing a node also implies in a higher probability of transmission, and therefore the observed results are as expected. 

To reinforce the validity of our Markov chain approach, we show that the agreement with the simulations also occurs at a node level. Fig.~\ref{fig:scatters}.b) compares the prevalence of each node as predicted by the numerical analysis (x axis) and as obtained from the simulations (y axis), using the same network sample and parameters of the Monte Carlo simulations. The black dashed line represents the identity line.

\begin{figure}
    \centering
    \includegraphics[trim=0 0 150 80, clip, scale=0.25]{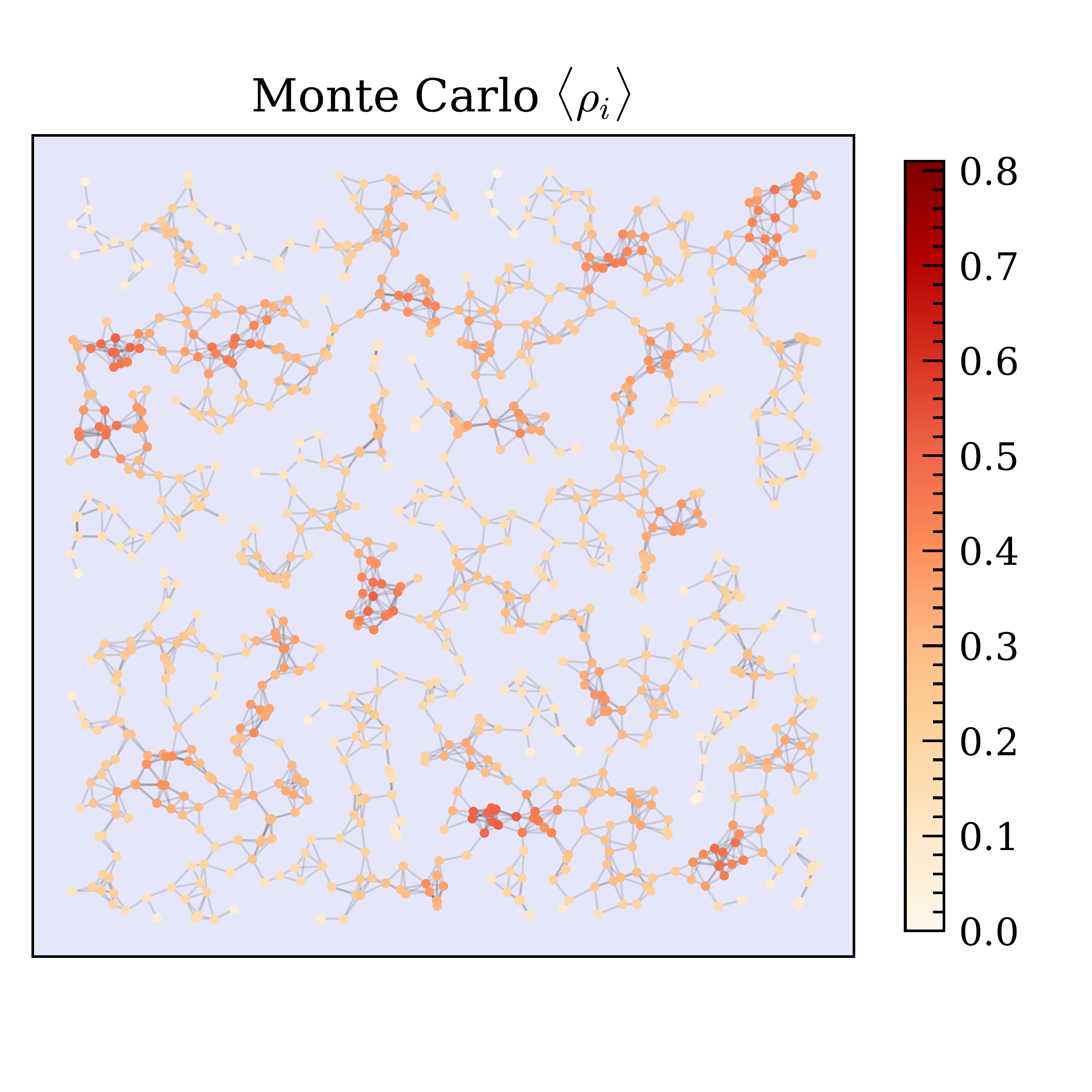}
    \includegraphics[trim=0 0 0 80, clip, scale=0.25]{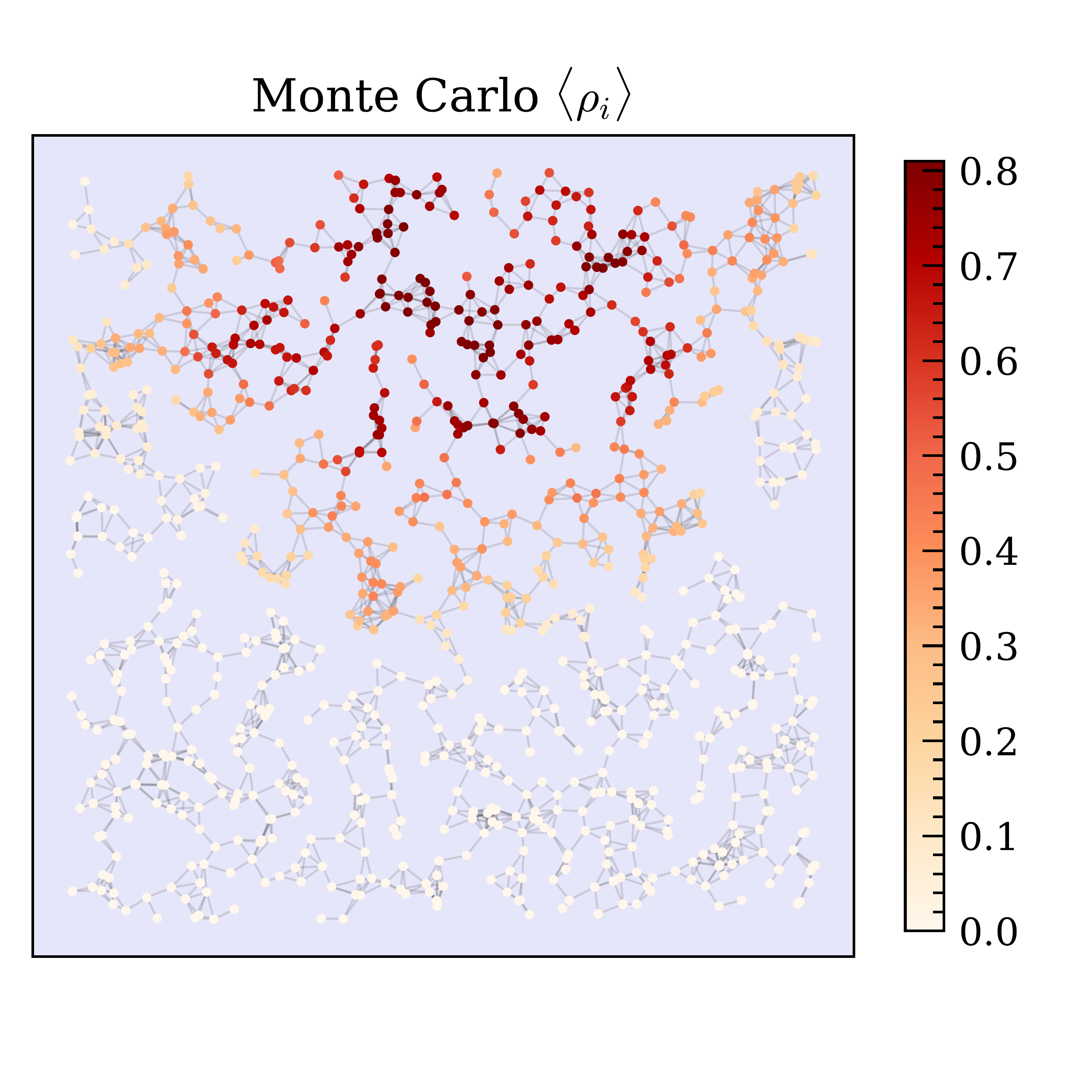}
    \caption{Average disease prevalence ($\rho_i$) in each node, without (left) and with (right) the attraction values, calculated from the same Monte Carlo simulations and network sample used for Fig.~\ref{fig:scatters}. The average prevalence for the whole system is of $0.64$ with the attraction and $0.44$ without it. }
    \label{fig:network_prevalence}
\end{figure}

Finally, we show a visual representation of the network and the average prevalences of each node, and compare with the case when no attraction landscape is present. The results are depicted in Fig.~\ref{fig:network_prevalence}, which uses the same Monte Carlo simulations as in Fig.~\ref{fig:scatters}. As the RGG topology is considerably homogeneous, the lack of an attraction landscape makes the overall prevalence (and node occupation) to be reasonably even across the space, as shown in Fig.~\ref{fig:network_prevalence} left. As we turn on the attraction (Fig.~\ref{fig:network_prevalence} right), individuals are driven to the surroundings of the Gaussian peak, and so is the disease prevalence. As already shown in Fig.~\ref{fig:gradexponent}, a higher attraction causes an overall higher disease prevalence. Indeed, in Fig.~\ref{fig:network_prevalence}, the average prevalence is $0.64$ for the network with attraction and $0.44$ without it. From the plots of Figs.~\ref{fig:gradexponent} and \ref{fig:network_prevalence}, we notice that the attraction landscape concentrate the agents in a smaller portion of the network, thus generating greater density of contacts and inducing greater disease transmission.

This shows that complexity, in this case produced by the heterogeneity of the attraction values, facilitates the disease propagation in metapopulations of small number of agents. This is already known for simple heterogeneous networks of contacts, in which nodes represent individuals \cite{pastor2001epidemic,barthelemy2005dynamical}, and for metapopulations in which nodes represent sites with high numbers of individuals \cite{colizza2007reaction}. In the mentioned works, the complexity comes from the network topology itself (for example, scale-free topologies), and not from an attraction landscape.

\section{Conclusion} \label{sec:conclusion}
In this work, we studied an SIS epidemic model in metapopulations of small number of agents, comparable to the number of nodes of the network. The mobility is defined by a networked random walk, and guided by an attraction landscape imposed over the nodes. Mobility and epidemic updates occur separately, with a parameter ($r$) controlling the proportion at which these updates occur. We develop a numerical dynamical process based on Markov chains to approximate the actual dynamics of the system, using an exact transition matrix to the epidemic step and an artifice to obtain approximate probability distributions for the mobility step. 

The numerical procedure was validated for different values of the epidemic-mobility ratio, transmission probability and attraction intensity, being considerably accurate to predict the stationary disease prevalence in the desired regime. This contrasts with the simpler procedure that considers only the average numbers, which is employed for larger populations but does not work for the contemplated regime. We study the effect of the attraction landscape by analysing the prevalence as a function of the intensity of the attraction values. The results show that, as expected, the propagation is greater when the attraction is sharper (i.e., when nodes are more heterogeneous), in agreement with what is known for complex networks of individual contacts~\cite{pastor2001epidemic} and metapopulations without attraction landscape~\cite{colizza2007invasion}. In addition, we analyse the epidemic propagation in more detail by plotting the local average prevalence for each node. We observe a representative local (node-wise) agreement between the Markov chain and the simulations.  It was also verified that the most visited nodes, regarded as hubs of the process, sustain a greater fraction of infectious individuals.

\blue{
As an approximation that goes one step beyond the basic mean field (see Fig. \ref{fig:epidemic_curves}), our model also has its domain of validity. For larger densities of agents $\eta$, the agreement of our model is expected to be improved,
but for $\eta \gg 1$ the computation becomes unfeasible, as the number of internal states of each node increases quadratically with this parameter. Alternatives that consider only the most likely states in each node could be used, in future works, possibly enhancing the precision with respect to the basic mean field using less computational complexity. On the other hand, for $\eta \ll 1$, the dynamics is driven only by a few states that have an appreciable probability. While this makes the computation faster, the agreement with Monte Carlo simulations is expected to decrease, and a more precise way to deal with the random walk would be required for better accuracy of the formulation. Therefore, the proposed model is most useful for modeling systems with a few individuals per site. 
}

We envision several extensions of this work, such as the consideration of other compartmental models, such as the classical SIR or SEIR; the study of the effect of the topology over the dynamics by considering other topological models of the graph. We only tested the numerical evaluation of the Markov chain using binomial distributions. However, and as shown in the appendix, any proposed distribution that maintains the average numbers of individuals is physically consistent, therefore other distributions -- such as the beta-binomial, more flexible than the regular binomial -- could be tested for better adjustment to the experimental values. Using other network topologies is another possible extension. For example, using a network that is more heterogeneous than the simple RGG, but can still be embedded in a 2D space, can provide further insights from the combination of network and landscape heterogeneity.

Finally, our work represents a step toward the description of epidemic models in an under-explored regime, providing new insights about this important problem. The proposed Markov chain approach could be applied to other systems for which exact solution is unfeasible, particularly for populations of small numbers of individuals.

\section{Acknowledgements}
\label{sec:ack}

P.C.V. acknowledges FAPESP for the research grant n. 2016/24555-0. E.K.T. acknowledges FAPESP grant n. 2019/01077-3. L.da F.C. thanks CNPq (grant 307085/2018-0) and FAPESP (grant 15/22308-2). 
F.A.R. acknowledges CNPq (grant 309266/2019-0) and FAPESP (grant 19/23293-0) for the financial support given for this research.
The authors also thank CAPES and the University of São Paulo.

\bibliographystyle{unsrt}
\bibliography{main}

\section*{Appendix}

Our proposed Markov chain approach is based on some approximations that reproduce the actual behavior of the system without the need of detailed calculations. It is essential, however, to show that it is able to keep physical consistency. In this appendix, we show that the probability distributions proposed to initialize the epidemic stage after a mobility update result in appropriate average quantities, including that of the total number of individuals (which is in fact constant during a simulation).

We first show that, by using Eqs. \ref{eq:binomial-poisson_pn} and \ref{eq:binomial-poisson_pm} for $p\big(\ninfrvar(t+1) = \ninfval, \; \popsize_{i}(t+1) = n_i \big )$ results in the actual average number of infectious agents after the mobility step, as expected. This average can be calculated as:

\begin{align}
\nonumber
    \average{\ninfrvar(t+1)} &= \sum_{n_i = 0}^{\infty} \sum_{\ninfval = 0}^{n_i} m_i \cdot p\big(\ninfrvar(t+1) = \ninfval, \; \popsize_{i}(t+1) = n_i \big )  
\\
    &= \sum_{n_i = 0}^{\infty} p(\popsize_i(t+1) = n_i) \sum_{\ninfval = 0}^{n_i} \ninfval \cdot p(\ninfrvar(t+1) = \ninfval | \popsize_i(t+1) = n_i),
    \label{eq:average_mi_open}
\end{align}

Where the second step breaks the joint probabilities into a product of a simple and a conditional probability, according to the multiplication rule.~Eq.~\ref{eq:binomial-poisson_pm} provides the value of the conditional one as a binomial in $\ninfval$, whose average value is $n_i b_i / a_i$. This is the only condition regarding $p(\ninfrvar(t+1) = \ninfval | \popsize_i(t+1) = n_i)$ that needed to be assume, so other distributions with the same condition can be used. Eq.~\ref{eq:average_mi_open} results into:

\begin{align}
    \average{\ninfrvar(t+1)} &= \sum_{n_i = 0}^{\infty} p(\popsize_i(t+1) = n_i) \; n_i \frac{b_i}{a_i}.
    \label{eq:average_mi_open-2}
\end{align}

Now, using the fact that $p(\popsize_i(t+1) = n_i)$, according to Eq.~\ref{eq:binomial-poisson_pn} (which refers to a Poison distribution), produces an average value of $\average{\popsize_i} = a_i$, the $a_i$ in the denominator of Eq. \ref{eq:average_mi_open-2} cancels that of average $n_i$, resulting that $\average{\ninfrvar(t+1)} = b_i$, as assumed by its definition (see Eq. \ref{eq:avg_next_ni_mobility}). Notice that, again, the only condition over the distribution of $p(\popsize_i(t+1) = n_i)$ is that its average is $a_i$.  Therefore, the same holds for the binomial approximation proposed in Eq. \ref{eq:binom_approx} and, again, other distributions could be tested for better results.

We now show that, although the local average occupations of each node ($\average{\popsize_i}$) are modified by the mobility step, the total average $\average{\popsize}$ of the number of agents is preserved. We calculate this number  in $t = 1$, using Eq. \ref{eq:avg_next_n_mobility} to get:

\begin{align}
    \average{\popsize(t+1)} &= \sum_{i=1}^{\numnodes} \average{\popsize_i(t+1)}
     \nonumber \\
    &= \sum_{i=1}^{\numnodes} \sum_{j=1}^{\numnodes} \average{\popsize_j(t)} \ptravel_{ji}
\end{align}

Where, for convenience, we set $\ptravel_{ij} = 0$ between nodes $i$ and $j$ that are not linked. Reorganizing the sum and using the stochastic property of matrix $\ptravel$, we get:

\begin{align}
    \average{\popsize(t+1)} &= \sum_{j=1}^{\numnodes} \average{\popsize_j(t)} \sum_{i=1}^{\numnodes} \ptravel_{ji} = \sum_{j=1}^{\numnodes} \average{\popsize_j(t)} = \average{N(t)},
\end{align}

As we wanted to demonstrate.

\end{document}